\documentclass[3p, 12pt, sort&compress]{elsarticle}
\usepackage{float} 
\usepackage{graphicx}
\usepackage{amsmath}
\usepackage{amssymb}
\usepackage{bm}
\usepackage{framed}
\usepackage{subfigure}
\usepackage{hyperref}
\usepackage{nomencl}
\usepackage{color}
\usepackage{textcomp}
\usepackage{graphics}
\usepackage{epsfig}
\usepackage{bm}
\usepackage{epstopdf}
\usepackage{amsthm}
\usepackage{float}
\usepackage{latexsym,amsfonts}
\usepackage{url}
\usepackage{longtable}
\usepackage[figuresright]{rotating}
\usepackage{listings}
\usepackage{algorithm}
\usepackage{algpseudocode}
\newlength\figureheight 
\newlength\figurewidth 
\usepackage{pgfplots}
\usepackage{scalefnt}
\usepackage{graphicx}
\usepackage{amsmath}
\usepackage{amssymb}
\usepackage{bbm}
\usepackage{bm}
\usepackage{framed}
\usepackage{multirow}
\usepackage{nomencl}
\usepackage{color}
\usepackage{textcomp}
\usepackage{graphics}
\usepackage{epsfig}
\usepackage{bm}
\usepackage{epstopdf}
\usepackage{courier}
\usepackage{amsthm}
\usepackage{float}
\usepackage{latexsym,amsfonts}
\usepackage{url}
\usepackage{longtable}
\usepackage[figuresright]{rotating}
\usepackage{listings}
\usepackage{algorithm}
\usepackage{algpseudocode}
\usepackage{flushend}
\usepackage{footnote}
\usepackage{tikz}
\usetikzlibrary{shapes}
\makesavenoteenv{tabular}

\theoremstyle{definition}

\makeatletter
\def\ps@pprintTitle{%
   \let\@oddhead\@empty
   \let\@evenhead\@empty
   \let\@oddfoot\@empty
   \let\@evenfoot\@oddfoot
}
\makeatother

\begin{document}

\begin{frontmatter}
\title{Efficient Estimation of the Left Tail of Bimodal Distributions with Applications to Underwater Optical Communication Systems}
\author[rvt]{Chaouki Ben Issaid}
\ead{chaouki.benissaid@kaust.edu.sa}

\author[rvt]{Mohamed-Slim Alouini}
\ead{slim.alouini@kaust.edu.sa}

\address[rvt]{King Abdullah University of Science and Technology (KAUST),\\
Computer, Electrical, and Mathematical Sciences and Engineering (CEMSE) Division,\\
Thuwal, Makkah Province, 23955-6900, Saudi Arabia}

\begin{abstract}
In this paper, we propose efficient importance sampling estimators to evaluate the outage probability of maximum ratio combining receivers over turbulence-induced fadings in underwater wireless optical channels. We consider two fading models: exponential-lognormal, and exponential-generalized Gamma. The cross-entropy optimization method is used to determine the optimal biased distribution. We show by simulations that the number of samples required by importance sampling estimator is much less compared to naive Monte Carlo for the same accuracy requirement.\\
\end{abstract}

\end{frontmatter}
\section{Introduction}
Nowadays, many of the underwater exploration and exploitation activities require the setting up of high-speed data transmission links. Traditional solutions like cable or fiber communication are costly and present very restricted flexibility. Alternatively, acoustic communications offer very low functioning performance. On the other hand, underwater wireless optical communication (UWOC) systems started to gain popularity since it offers a more flexible and higher-speed transmission solution compared to the aforementioned techniques \cite{ZA}-\cite{HZI}. However, the installation of UWOC systems can be challenging. In fact, the optical signal is severely affected by both absorption and scattering \cite{DL} in addition to the fading when operating in a turbulent environment \cite{JU, XU}.

Recently, some statistical models to describe the turbulence faced by UWOC channels have been proposed. Inspired from the classical lognormal turbulence model used in free space optical communication systems, some papers, such as \cite{LiuS} and \cite{YiU}, have considered the lognormal distribution to model the irradiance fluctuations in the underwater environment. However, not only this kind of model has not been confirmed by experimental measurements, the structure of the refractive-index spectrum in the atmosphere is different than the one in the water \cite{ZAZ}. In \cite{MS}, the authors proposed a bimodal exponential-lognormal distribution to model the turbulence. They showed that the proposed model fits the experimental distribution when the scintillation index is between the values $0.1$ and $1$. In \cite{EA}, Zedini \textit{et al.} presented a more robust model based on a bimodal exponential-Gamma distribution. The authors used the expectation maximization algorithm to determine the maximum likelihood parameter for the new model. Numerical simulations show that the proposed model provide a perfect fit with the collected data under different turbulence conditions. A more general model has been proposed in the journal version of \cite{EA} where the authors generalized the model discussed in \cite{EA} by taking a bimodal exponential-generalized Gamma fading model.

Thanks to different diversity techniques \cite[Chap. 9]{DMM}, we can combine the signals in order to reduce the fading effect. There are more or less complex linear combination techniques which make it possible to recover a signal with a good average level, in particular, we find the maximum ratio combining (MRC) technique. Finding exact outage probability of multibranch MRC diversity receivers over the aforementioned fading channels seems to be quite challenging. In this case, this performance metric can be estimated by means of numerical methods, for instance naive Monte Carlo (MC) method. In this paper, we propose an efficient importance sampling (IS) based simulation method in order to reduce the number of simulation runs required especially when low outage probabilities requirement are needed.

The reminder of this paper is organized as follows. The system model is described in Section \ref{section2}. We then recall briefly the main idea behind IS method in Section \ref{section3}. In Section \ref{section4}, we introduce the cross-entropy (CE) method and we present the proposed approach to estimate the outage probability in our particular set-up. In Section \ref{section5}, some selected numerical results are presented to show the significant improvement that the proposed approach offers compared to naive MC.

Throughout this paper, the following notations are used: $L$ denotes the number of diversity branches, $\mathbf{X} = \left(X_1,\dots, X_L \right)$ is a random vector, where $\{X_\ell\}_{\ell=1}^{L}$ represent the fading powers, and their sum is $S_L(\mathbf{X}) = \sum\limits_{\ell=1}^{L}{X_\ell}$. The probability density function (PDF) of $X_{\ell}$ is $f_{X_{\ell}}(\cdot)$, and the joint PDF is given by $f(\mathbf{x}) = \prod\limits_{\ell=1}^{L}{f_{X_{\ell}}(x)}$.
\section{System Model}\label{section2}
The expression of the instantaneous signal-to-noise ratio (SNR) expression at the MRC receiver is defined as \cite{AFM}
\begin{align}
\gamma_{end} = \frac{E_s}{N_0} \sum\limits_{\ell=1}^{L}{X_\ell},
\end{align}
where $\frac{E_s}{N_0}$ is the SNR per symbol. The outage probability, in this case, is defined as
\begin{align}\label{outp}
P = \mathbb{P}(\gamma_{end} \leq \gamma_{th}) = \mathbb{P}( S_L(\mathbf{X}) \leq \gamma_0), 
\end{align}
where $\gamma_0 = \frac{N_0}{E_s} \gamma_{th}$ and $\gamma_{th}$ is a given threshold.

In the rest of the paper, we assume that $\{X_\ell\}_{\ell=1}^{L}$ are independent and that the PDF of $X_{\ell}$, $f_{X_{\ell}}(\cdot)$, is a bimodal distribution having one of these two forms
\begin{itemize}
\item Exponential-lognormal model (EXP-LN) \cite{MS}\\
$f_{X_{\ell}}(x) = \frac{\omega}{\lambda_\ell} \exp\left(-\frac{x}{\lambda_\ell}\right) + \frac{1-\omega}{x \sigma_\ell \sqrt{2 \pi}} \exp\left(-\frac{(\log(x)-\mu_\ell)^2}{2 \sigma_\ell^2}\right)$
\item Exponential-generalized Gamma model (EXP-GG) \cite{EA} \\
$f_{X_{\ell}}(x) = \frac{\omega}{\lambda_\ell} \exp\left(-\frac{x}{\lambda_\ell}\right) + \frac{(1-\omega) \beta_\ell \alpha_\ell^{\alpha_\ell}}{\Gamma(\alpha_\ell) \Omega^\alpha_\ell} \exp\left(-\frac{\alpha_\ell}{\Omega_\ell} x^{\beta_\ell} \right)$
\end{itemize}
where $\omega \in [0,1]$ is a weighting parameter and $\lambda_\ell$, $\mu_\ell$, $\sigma_\ell$, $\alpha_\ell$, $\beta_\ell$, and $\Omega_\ell$ are positive real numbers that represent the parameters of each distribution $f_{X_{\ell}}(\cdot)$.  
\section{Importance Sampling}\label{section3}
The IS method \cite{IJ} is the most used method when dealing with the estimation of very small probabilities. The aim is to reduce the variance of the naive MC estimator by introducing a new biased PDF. The implementation of the method is straightforward, however the gain in terms of number of simulation runs (equivalently the reduction in the variance) highly depends on the choice of the biased PDF. Choosing an optimal biased PDF is not a trivial task quite often and represents the corner stone of the proposed IS scheme. The IS estimator of (\ref{outp}) is given by
\begin{align}
\hat{P}_{IS} = \frac{1}{N} \sum_{i=1}^{N}{\mathbbm{1}_{\left( S_L(\mathbf{X}(\omega_i)) \leq \gamma_0 \right)} \frac{f(\mathbf{X}(\omega_i))}{f^*(\mathbf{X}(\omega_i))}},
\end{align}
where $\mathbbm{1}_{(\cdot)}$ is the indicator function and $\{X_{\ell}(\omega_i)\}_{\ell=1}^{L}$ are independent samples from 
\begin{align}
f^*(\mathbf{x}) = \prod_{\ell=1}^{L}{f^{*}_{X_{\ell}}(x)}.
\end{align}
The IS estimator variance can be written as
\begin{align}
\mathbb{V}^*\left[\hat{P}_{IS}\right] = \frac{1}{N} \left(\mathbb{E}^*\left[\mathbbm{1}_{\left( S_L(\mathbf{X}) \leq \gamma_0 \right)} \left(\frac{f(\mathbf{X})}{f^*(\mathbf{X})}\right)^2\right] - P^2\right),
\end{align}
where $\mathbb{V}^*[\cdot]$ and $\mathbb{E}^*[\cdot]$ are respectively the variance and the expected value w.r.t the biased PDFs.

The variance of the IS estimator depends in particular on the choice of the biased PDF $f^{*}(\cdot)$. If it is well chosen, then the variance of the IS estimator can become very low. Otherwise, the variance of the IS estimator may even be much higher than the naive MC estimator. Since the variance is a non-negative quantity, ideally we would like to choose $f^{*}(\cdot)$ such that the variance of the IS estimator is zero, that is \cite[Chap. 4]{IJ}
\begin{align}
\mathbb{E}^*\left[\mathbbm{1}_{\left( S_L(\mathbf{X}) \leq \gamma_0 \right)} \left(\frac{f(\mathbf{X})}{f^*(\mathbf{X})}\right)^2\right] = P^2.
\end{align}
It is well known that the optimal solution in this case is given by \cite[Chap. 4]{IJ}
\begin{align}\label{opt}
f^{*}_{opt}(\mathbf{x}) = \frac{\mathbbm{1}_{\left( S_L(\mathbf{x}(\omega_i)) \leq \gamma_0 \right)} f(\mathbf{x})}{P}.
\end{align}
The biased PDF $f^{*}_{opt}(\cdot)$ unfortunately depends on the probability to be estimated $P$. The PDF $f^{*}_{opt}(\cdot)$ cannot be used in practice. It is necessary to perform an optimization in order to determine a biased PDF $f^{*}(\cdot)$ such that it approximates well the density $f^{*}_{opt}(\cdot)$. There are several methods in the literature to determine a good approximation for the optimal biased density. In this paper, we chose to combine IS with the Cross-Entropy (CE) method. The advantage of such approach is (i) it does not require further conditions unlike conventional IS approaches, for instance the existence of the moment generating function for exponential twisting IS, and (ii) the optimal biased density can be obtained using standard optimization techniques. 
\section{Cross-Entropy}\label{section4}
The CE method \cite{GB, RH, TR} allows to approximate the optimal biased density of the IS scheme among a family of parametric PDFs w.r.t the crossed entropy criterion. The crossed entropy, also called the Kullback-Leibler distance $\mathcal{D}(q, p)$ between two probability densities $p$ and $q$ is defined as \cite[Chap. 1]{TR}
\begin{align}\label{ce}
\nonumber \mathcal{D}(q, p) &= \mathbb{E}_{q}\left[\log\left(\frac{q(x)}{p(x)}\right)\right] \\
&= \int{q(x) \log(q(x)) dx} - \int{q(x) \log(p(x)) dx},
\end{align}
where $\mathbb{E}_q[\cdot]$ is the expected value w.r.t the density $q(\cdot)$.

The Kullback-Leibler metric is not really a distance in the mathematical sense of the term, but it allows to establish a criterion of deviation between two probability densities. In our case, these two probability densities are the optimal biased density $f^{*}_{opt}(\cdot)$ and a parametric biased density $f^{*}_{\nu}(\cdot)$. So, CE aims to minimize the crossed entropy between $f^{*}_{opt}(\cdot)$ and $f^{*}_{\nu}(\cdot)$. Since the first integral in (\ref{ce}) is independent of $f^{*}_{\nu}(\cdot)$, CE focus on minimizing the second term w.r.t the parameter $\nu$, that is to choose the best approximation among the family of distribution indexed by the parameter $\nu$, which turns to be equivalent to the following maximization problem \cite{TR}
\begin{align}\label{max}
\underset{\nu}\max{\int{f^{*}_{opt}(\mathbf{x}) \log(f^{*}_{\nu}(\mathbf{x})) d\mathbf{x}}}
\end{align}
Replacing (\ref{opt}) in (\ref{max}), the optimization problem is equivalent to 
\begin{align}
\underset{\nu}\max{}~\mathbb{E}\left[\mathbbm{1}_{\left( S_L(\mathbf{x}) \leq \gamma_0 \right)} \log(f^{*}_{\nu}(\mathbf{x}))\right].
\end{align}
Let $u$ be a reference parameter and the likelihood ratio $\mathcal{L}(\mathbf{x},u) = \frac{f(\mathbf{x})}{f^{*}_{u}(\mathbf{x})}$. The optimization problem can be re-written as
\begin{align}
\underset{\nu} \max{}~\mathbb{E}_u \left[\mathbbm{1}_{\left( S_L(\mathbf{x}) \leq \gamma_0 \right)} \mathcal{L}(\mathbf{x},u) \log(f^{*}_{\nu}(\mathbf{x}))\right],
\end{align}
and its stochastic counterpart is given by \cite{TR}
\begin{align}
\underset{\hat{\nu}} \max{}~\frac{1}{N} \sum_{i=1}^{N}{\mathbbm{1}_{\left( S_L(\mathbf{X}_i) \leq \gamma_0 \right)} \mathcal{L}(\mathbf{X}_i,u) \log(f^{*}_{\hat{\nu}}(\mathbf{X}_i))}.
\end{align}
Generally speaking, the above function is convex and differentiable w.r.t $\hat{\nu}$ \cite{RD}, thus the optimal solution is given by solving
\begin{align}\label{eqq}
\sum_{i=1}^{N}{\mathbbm{1}_{\left( S_L(\mathbf{X}_i) \leq \gamma_0 \right)} \mathcal{L}(\mathbf{X}_i,u) \nabla \log(f^{*}_{\hat{\nu}}(\mathbf{X}_i))} = 0,
\end{align}
where $\{\mathbf{X}_i\}_{i=1}^{N}$ are sampled from $f^{*}_{u}(\mathbf{x})$.

In our setting, we will choose the marginal density to be $f^{*}_{\nu_\ell}(x) = \frac{1}{\nu_\ell} \exp\left(-\frac{x}{\nu_\ell}\right)$, for $\ell=1,\dots, L$. This choice is justified by (i) the exponential distribution present a heavier left tail compared to the lognormal and generalized-Gamma distributions and (ii) with this particular choice, we can easily determine a closed-form  expression for the parameter of the biased distribution. In fact, we can show that, with this particular choice, (\ref{eqq}) leads to
\begin{align}\label{update}
\hat{\nu}_\ell = \frac{\sum_{i=1}^{N}{\mathbbm{1}_{\left( S_L(\mathbf{X}_i) \leq \gamma_0 \right)} \mathcal{L}(\mathbf{X}_i,u)} \mathbf{X}_{i\ell}}{\sum_{i=1}^{N}{\mathbbm{1}_{\left( S_L(\mathbf{X}_i) \leq \gamma_0 \right)} \mathcal{L}(\mathbf{X}_i,u)}},~\ell=1,\dots,L.
\end{align}
We provide the CE algorithm to compute the IS estimator based on \citep[Chap. 3]{TR} in Algorithm \ref{algo1}.
\begin{algorithm}[H]
  \caption{CE Algorithm}\label{algo1}
  \begin{algorithmic}[1]    
      \State Define $\hat{\nu}_0 = \lambda$ and set $t=1$.
      \State Generate $\{\mathbf{X}_i\}_{i=1}^{N}$ from $f^{*}_{\hat{\nu}_{t-1}}(\cdot)$.
      \State Compute  $\hat{\gamma}_t = S_{\lceil(1-\rho) N\rceil}$. 
      \State If $\hat{\gamma}_t < \gamma_0$, set $\hat{\gamma}_t = \gamma_0$.
      \State Use the same sample $\{\mathbf{X}_i\}_{i=1}^{N}$ to compute the updating formula (\ref{update}) with $u=\hat{\nu}_{t-1}$.
      \State If $\hat{\gamma}_t > \gamma_0$, set $t=t+1$ and go to step 2, otherwise go to step 7.
      \State Compute the IS estimator  given by
      \begin{align}
      \hat{P}_{IS} = \frac{1}{N} \sum_{i=1}^{N}{\mathbbm{1}_{\left( S_L(\mathbf{X}(\omega_i)) \leq \gamma_0 \right)} \frac{f(\mathbf{X})}{f^*_{\hat{\nu}_{t}}(\mathbf{X})}}.
      \end{align}
  \end{algorithmic}
\end{algorithm}
In practice, the number of samples used in step 2 to determine the optimal parameter can be less than the one used in step 7 to compute the IS estimator. The parameter $\rho$ is used to compute adaptively $\hat{\gamma}_t = S_{\lceil(1-\rho) N\rceil}$, an estimator of $\gamma_t$, the sample $(1-\rho)$ quantile of $S_L(\mathbf{X})$ under $\hat{\nu}_{t-1}$, where $\lceil x\rceil$ denotes is the smallest integer greater than or equal to $x$.

\section{Numerical Simulations}\label{section5}

\begin{table*}[t]
\centering
\caption{Simulation parameters used to simulate the outage probability of $L$-branch MRC diversity receivers.}
\label{tab}
\begin{tabular}{|c|c|c|}
\hline
\multirow{2}{*}{Model} & \multicolumn{2}{c|}{Simulation parameters} \\ \cline{2-3} 
                       & $L=2$                & $L=4$               \\ \hline
                       &                      &                      \\
EXP-LN                 & $\omega = 0.2045, \lambda_1 = 0.5389, \lambda_2 = 0.9786$  & $\omega=0.2045, \lambda_1= 0.5389, \lambda_2= 0.9786$ \\
                       & $ \sigma_1=\sigma_2=0.0253, \mu_1=\mu_2=0.1117$ & $ \lambda_3 = 0.4854, \lambda_4 = 0.224$ \\
                       & & $\sigma_1=\sigma_2=\sigma_3=\sigma_4=0.0253$ \\ 
                       &  & $\mu_1=\mu_2=\mu_3=\mu_4=0.1117$ \\ \hline
                       &                      &                      \\
EXP-GG                 & $\omega = 0.4876, \lambda_1 = 0.5389, \lambda_2 = 0.9786$  & $\omega=0.4876, \lambda_1= 0.5389, \lambda_2= 0.9786$     \\ 
& & $\lambda_3 = 0.4854, \lambda_4 = 0.224$ \\
& $\alpha_1=\alpha_2 = 3.275, \beta_1= \beta_2= 1.45$ & $\alpha_1=\alpha_2 = \alpha_3 = \alpha_4 = 3.275$ \\
& &  $\beta_1= \beta_2= \beta_3 = \beta_4 = 1.45$\\ \hline
\end{tabular}
\end{table*}

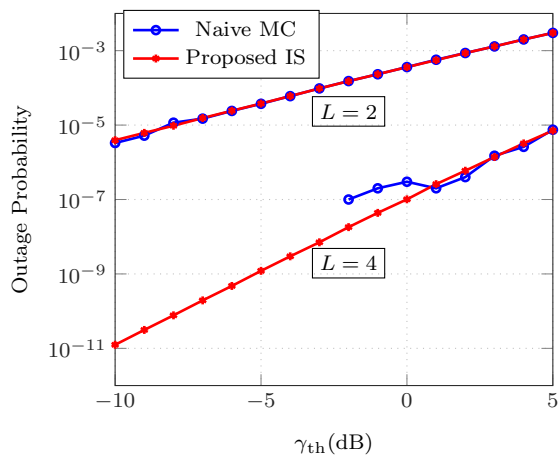
\begin{figure}[H]
\centering
\setlength\figureheight{0.3\textwidth}
\setlength\figurewidth{0.35\textwidth}
\scalefont{0.7}
\begin{tikzpicture}

\begin{semilogyaxis}[%
width=\figurewidth,
height=\figureheight,
at={(0\figurewidth,0\figureheight)},
scale only axis,
every outer x axis line/.append style={darkgray!60!black},
every x tick label/.append style={font=\color{darkgray!60!black}},
xmin=-10,
xmax=5,
xlabel={$\gamma{}_{\text{th}}\text{(dB)}$},
xmajorgrids,
every outer y axis line/.append style={darkgray!60!black},
every y tick label/.append style={font=\color{darkgray!60!black}},
ymin=1e-12,
ymax=0.01,
yminorticks=true,
ymajorgrids,
yminorgrids,
grid style={dotted},
ylabel={Outage Probability},
legend style={at={(0.018574218750001,0.824593784825617)},anchor=south west,draw=black,fill=white,align=left}
]
\addplot [
color=blue,
solid,
line width=1.0pt,
mark size=1.5pt,
mark=o,
mark options={solid},
]
  table[row sep=crcr]{%
-10	3.3e-06\\
-9	5.2e-06\\
-8	1.16e-05\\
-7	1.49e-05\\
-6	2.4e-05\\
-5	3.73e-05\\
-4	5.99e-05\\
-3	9.66e-05\\
-2	0.0001521\\
-1	0.0002321\\
0	0.0003623\\
1	0.0005694\\
2	0.0008644\\
3	0.0013\\
4	0.002\\
5	0.003\\
};
\addlegendentry{Naive MC};
\addplot [
color=red,
solid,
line width=1.0pt,
mark size=1.5pt,
mark=asterisk,
mark options={solid},
]
  table[row sep=crcr]{%
-10	3.9186e-06\\
-9	6.1736e-06\\
-8	9.6019e-06\\
-7	1.5451e-05\\
-6	2.4251e-05\\
-5	3.7893e-05\\
-4	6.0131e-05\\
-3	9.624e-05\\
-2	0.00015133\\
-1	0.00023317\\
0	0.00036059\\
1	0.00055675\\
2	0.00086125\\
3	0.0013\\
4	0.002\\
5	0.003\\
};
\addlegendentry{Proposed IS};
\addplot [
color=blue,
solid,
line width=1.0pt,
mark size=1.5pt,
mark=o,
mark options={solid},
]
  table[row sep=crcr]{%
-10	0\\
-9	0\\
-8	0\\
-7	0\\
-6	0\\
-5	0\\
-4	0\\
-3	0\\
-2	1e-07\\
-1	2e-07\\
0	3e-07\\
1	2e-07\\
2	4e-07\\
3	1.5e-06\\
4	2.6e-06\\
5	7.5e-06\\
};
\addplot [
color=red,
solid,
line width=1.0pt,
mark size=1.5pt,
mark=asterisk,
mark options={solid},
]
  table[row sep=crcr]{%
-10	1.2455e-11\\
-9	3.1279e-11\\
-8	7.7107e-11\\
-7	1.9449e-10\\
-6	4.752e-10\\
-5	1.2093e-09\\
-4	2.9733e-09\\
-3	7.0748e-09\\
-2	1.8085e-08\\
-1	4.3731e-08\\
0	1.0104e-07\\
1	2.5864e-07\\
2	5.9304e-07\\
3	1.3899e-06\\
4	3.2153e-06\\
5	7.1121e-06\\
};
\node[black,draw,below] at (axis cs:-2,6.0e-05){{$L = 2$}};
\node[black,draw,below] at (axis cs:-2,5.0e-09){{$L = 4$}};
\end{semilogyaxis}
\end{tikzpicture}%
\caption{Outage probability of $L$-branch MRC diversity receivers over exponential-lognormal fading model with $E_s/N_0=10$ dB. Number of samples $N = 10^7$ and $N^{*} = 10^4$.}
\label{fig01}
\end{figure}

Table \ref{tab} summarizes the parameters used in the simulation part of this paper based on \cite{EA}. For the computation of $\hat{\gamma}_t$ in Algoritm \ref{algo1}, we used the value $\rho=0.01$ \cite{TR}. We plot in Fig. \ref{fig01} (respectively Fig. \ref{fig05}) the estimated outage probability using naive MC (in blue) and the proposed IS scheme (in red) against the threshold $\gamma_{th}$, for two cases $L \in \{2, 4\}$, over the exponential-lognormal (respectively the exponential-generalized Gamma) fading model. The number of samples used by naive MC in Fig. \ref{fig01} is $N=10^7$ which is greater than the one used by IS, $N^*=10^4$. However, the proposed IS scheme can accurately estimate outage probability up to $10^{-11}$ when $L=4$ unlike naive MC which fails to estimate outage probability lower than $10^{-6}$ unless more samples are taken. Similar conclusions can be drawn in the case of Fig. \ref{fig05}.

\begin{figure}[H]
\centering
\setlength\figureheight{0.3\textwidth}
\setlength\figurewidth{0.35\textwidth}
\scalefont{0.7}
\begin{tikzpicture}
    \begin{semilogyaxis}[
    width=\figurewidth,
    height=\figureheight,
    at={(0\figurewidth,0\figureheight)},
    scale only axis,
    every outer x axis line/.append style={darkgray!60!black},
    every x tick label/.append style={font=\color{darkgray!60!black}},
    xmin=-10,
    xmax=5,
    xlabel={$\gamma{}_{\text{th}}\text{(dB)}$},
    xmajorgrids,
    every outer y axis line/.append style={darkgray!60!black},
    every y tick label/.append style={font=\color{darkgray!60!black}},
    ymin=100,
    ymax=1e+16,
    ymajorgrids,
    yminorgrids,
    grid style={dotted},
    ylabel={Simulation Runs},
    legend entries={$L=2$, $L=4$}
    ]
    \addlegendimage{black, solid, no marks}
    \addlegendimage{black, dashed,no marks}
    \addplot [
    color=blue,
    solid,
    line width=1.0pt,
    mark size=1.5pt,  
    mark=o,
    mark options={solid},
    ]
    table[row sep=crcr]{%
    -10	392140000\\
    -9	248900000\\
    -8	160030000\\
    -7	99452000\\
    -6	63363000\\
    -5	40550000\\
    -4	25553000\\
    -3	15965000\\
    -2	10153000\\
    -1	6588800\\
     0	4259900\\
     1	2758500\\
     2	1982700\\
     3	993660\\
     4	783660\\
     5	520370\\
     };
     \addplot [
color=red,
solid,
line width=1.0pt,
mark size=1.5pt,
mark=asterisk,
mark options={solid},
]
  table[row sep=crcr]{%
-10	1598.5\\
-9	1595.9\\
-8	1586\\
-7	1594.7\\
-6	1552.4\\
-5	1551\\
-4	1595.9\\
-3	1557.6\\
-2	1558\\
-1	1504.9\\
0	1494.8\\
1	1447.8\\
2	1470.7\\
3	1434.2\\
4	1337.8\\
5	1294.6\\
};
\addplot [
color=blue,
dashed,
line width=1.0pt,
mark size=1.5pt,
mark=o,
mark options={solid},
forget plot
]
  table[row sep=crcr]{%
-10	123380000000000\\
-9	49126000000000\\
-8	19929000000000\\
-7	7900700000000\\
-6	3233700000000\\
-5	1270700000000\\
-4	516810000000\\
-3	217200000000\\
-2	84968000000\\
-1	34896000000\\
0	14394000000\\
1	5941300000\\
2	2105600000\\
3	857920000\\
4	477920000\\
5	216060000\\
};
\addplot [
color=red,
dashed,
line width=1.0pt,
mark size=1.5pt,
mark=asterisk,
mark options={solid},
forget plot
]
  table[row sep=crcr]{%
-10	2653\\
-9	2520.3\\
-8	2575.5\\
-7	2621.9\\
-6	2535.6\\
-5	2535.6\\
-4	2595.5\\
-3	2585.4\\
-2	2545.6\\
-1	2515.1\\
0	2452.5\\
1	2423.6\\
2	2402.9\\
3	2323.6\\
4	2260.1\\
5	2160.3\\
};
\node[black,draw,below] at (axis cs:-5.5,5.0e09){{Naive MC}};
\node[black,draw,below] at (axis cs:-0.5,8.0e04){{Proposed IS}};
\draw (axis cs:-2.5,2.0e03) ellipse (0.1cm and 0.15cm);
\draw (axis cs:-2.5,2.0e09) ellipse (0.2cm and 0.95cm);
    \end{semilogyaxis}
\end{tikzpicture}
\caption{Number of required simulation runs for $5\%$ relative error for $L$-branch MRC diversity receivers over exponential-lognormal fading model with $E_s/N_0=10$ dB. Solid line: $L=2$ and dashed line: $L=4$.}
\label{fig02}
\end{figure}
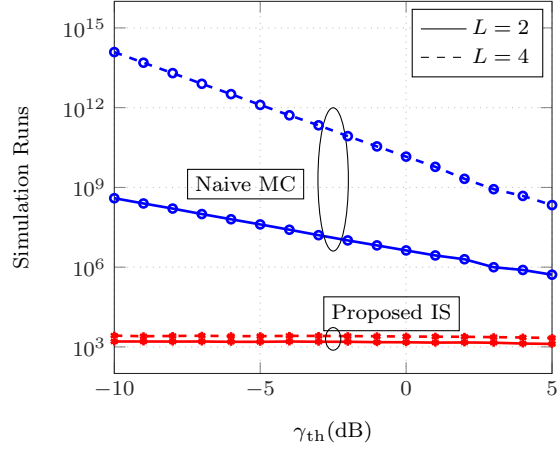

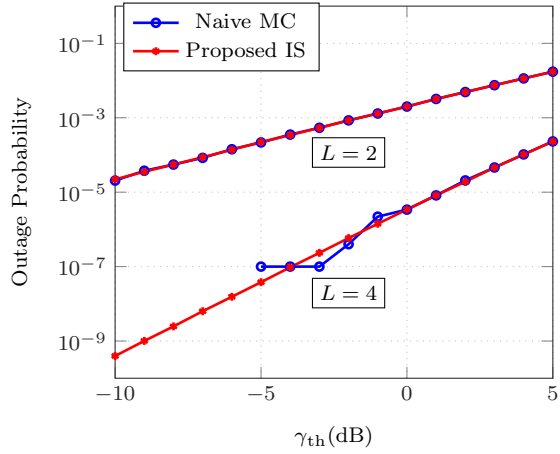
\begin{figure}[H]
\centering
\setlength\figureheight{0.3\textwidth}
\setlength\figurewidth{0.35\textwidth}
\scalefont{0.7}
\begin{tikzpicture}

\begin{semilogyaxis}[%
width=\figurewidth,
height=\figureheight,
at={(0\figurewidth,0\figureheight)},
scale only axis,
every outer x axis line/.append style={darkgray!60!black},
every x tick label/.append style={font=\color{darkgray!60!black}},
xmin=-10,
xmax=5,
xlabel={$\gamma{}_{\text{th}}\text{(dB)}$},
xmajorgrids,
every outer y axis line/.append style={darkgray!60!black},
every y tick label/.append style={font=\color{darkgray!60!black}},
ymin=1e-10,
ymax=1,
yminorticks=true,
ymajorgrids,
yminorgrids,
grid style={dotted},
ylabel={Outage Probability},
legend style={at={(0.018574218750001,0.824593784825617)},anchor=south west,draw=black,fill=white,align=left}
]
\addplot [
color=blue,
solid,
line width=1.0pt,
mark size=1.5pt,
mark=o,
mark options={solid},
]
  table[row sep=crcr]{%
-10	2.05e-05\\
-9	3.77e-05\\
-8	5.52e-05\\
-7	8.36e-05\\
-6	0.0001423\\
-5	0.0002169\\
-4	0.0003528\\
-3	0.0005383\\
-2	0.00085\\
-1	0.0013\\
0	0.002\\
1	0.0032\\
2	0.0049\\
3	0.0075\\
4	0.0114\\
5	0.0173\\
};
\addlegendentry{Naive MC};
\addplot [
color=red,
solid,
line width=1.0pt,
mark size=1.5pt,
mark=asterisk,
mark options={solid},
]
  table[row sep=crcr]{%
-10	2.2133e-05\\
-9	3.5455e-05\\
-8	5.5212e-05\\
-7	8.6799e-05\\
-6	0.00013839\\
-5	0.00022242\\
-4	0.00034323\\
-3	0.00053408\\
-2	0.00084721\\
-1	0.0013\\
0	0.002\\
1	0.0032\\
2	0.0049\\
3	0.0075\\
4	0.0114\\
5	0.0173\\
};
\addlegendentry{Proposed IS};
\addplot [
color=blue,
solid,
line width=1.0pt,
mark size=1.5pt,
mark=o,
mark options={solid},
]
  table[row sep=crcr]{%
-10	0\\
-9	0\\
-8	0\\
-7	0\\
-6	0\\
-5	1e-07\\
-4	1e-07\\
-3	1e-07\\
-2	4e-07\\
-1	2.2e-06\\
0	3.4e-06\\
1	8.2e-06\\
2	2.07e-05\\
3	4.6065e-05\\
4	0.00010347\\
5	0.00023122\\
};
\addplot [
color=red,
solid,
line width=1.0pt,
mark size=1.5pt,
mark=asterisk,
mark options={solid},
]
  table[row sep=crcr]{%
-10	3.9635e-10\\
-9	1.0074e-09\\
-8	2.4697e-09\\
-7	6.3253e-09\\
-6	1.5452e-08\\
-5	3.8163e-08\\
-4	9.594e-08\\
-3	2.3418e-07\\
-2	5.8336e-07\\
-1	1.4148e-06\\
0	3.4285e-06\\
1	8.2285e-06\\
2	1.9065e-05\\
3	4.6065e-05\\
4	0.00010347\\
5	0.00023122\\
};
\node[black,draw,below] at (axis cs:-2,3.0e-04){{$L = 2$}};
\node[black,draw,below] at (axis cs:-2,5.0e-08){{$L = 4$}};
\end{semilogyaxis}
\end{tikzpicture}%
\caption{Outage probability of $L$-branch MRC diversity receivers over exponential-generalized Gamma fading model with $E_s/N_0=10$ dB. Number of samples $N = 10^7$ and $N^{*} = 10^4$.}
\label{fig05}
\end{figure}

\begin{figure}[H]
\centering
\setlength\figureheight{0.3\textwidth}
\setlength\figurewidth{0.35\textwidth}
\scalefont{0.7}
\begin{tikzpicture}
    \begin{semilogyaxis}[
    width=\figurewidth,
    height=\figureheight,
    at={(0\figurewidth,0\figureheight)},
    scale only axis,
    every outer x axis line/.append style={darkgray!60!black},
    every x tick label/.append style={font=\color{darkgray!60!black}},
    xmin=-10,
    xmax=5,
    xlabel={$\gamma{}_{\text{th}}\text{(dB)}$},
    xmajorgrids,
    every outer y axis line/.append style={darkgray!60!black},
    every y tick label/.append style={font=\color{darkgray!60!black}},
    ymin=100,
    ymax=1e+16,
    ymajorgrids,
    yminorgrids,
    grid style={dotted},
    ylabel={Simulation Runs},
    legend entries={$L=2$, $L=4$}
    ]
    \addlegendimage{black, solid, no marks}
    \addlegendimage{black, dashed,no marks}
 \addplot [
color=blue,
solid,
line width=1.0pt,
mark size=1.5pt,
mark=o,
mark options={solid},
]
  table[row sep=crcr]{%
-10	69427000\\
-9	43339000\\
-8	27830000\\
-7	17702000\\
-6	11102000\\
-5	6907300\\
-4	4475400\\
-3	2875600\\
-2	1812200\\
-1	1159900\\
0	745940\\
1	483120\\
2	312720\\
3	203480\\
4	133070\\
5	87488\\
};
\addplot [
color=red,
solid,
line width=1.0pt,
mark size=1.5pt,
mark=asterisk,
mark options={solid},
]
  table[row sep=crcr]{%
-10	1580.3\\
-9	1564.3\\
-8	1586.3\\
-7	1602.3\\
-6	1566.5\\
-5	1529.2\\
-4	1548.4\\
-3	1548.2\\
-2	1532.9\\
-1	1529.8\\
0	1476.6\\
1	1492.2\\
2	1422.4\\
3	1403.1\\
4	1387.9\\
5	1372.3\\
};
\addplot [
color=blue,
dashed,
line width=1.0pt,
mark size=1.5pt,
mark=o,
mark options={solid},
forget plot
]
  table[row sep=crcr]{%
-10	3877000000000\\
-9	1525400000000\\
-8	622200000000\\
-7	242940000000\\
-6	99446000000\\
-5	40265000000\\
-4	16017000000\\
-3	6561800000\\
-2	2634100000\\
-1	1086100000\\
0	448200000\\
1	186740000\\
2	80597000\\
3	33357000\\
4	14849000\\
5	6644400\\
};
\addplot [
color=red,
dashed,
line width=1.0pt,
mark size=1.5pt,
mark=asterisk,
mark options={solid},
forget plot
]
  table[row sep=crcr]{%
-10	2635.6\\
-9	2610\\
-8	2608.1\\
-7	2574.5\\
-6	2606.3\\
-5	2575\\
-4	2545.7\\
-3	2544.3\\
-2	2483.6\\
-1	2510\\
0	2415.7\\
1	2425.7\\
2	2380.8\\
3	2297.6\\
4	2250.8\\
5	2236.6\\
};
\node[black,draw,below] at (axis cs:-5.5,5.0e09){{Naive MC}};
\node[black,draw,below] at (axis cs:-0.5,8.0e04){{Proposed IS}};
\draw (axis cs:-2.5,2.0e03) ellipse (0.1cm and 0.15cm);
\draw (axis cs:-2.5,9.0e07) ellipse (0.2cm and 0.95cm);
    \end{semilogyaxis}
\end{tikzpicture}
\caption{Number of required simulation runs for $5\%$ relative error for $L$-branch MRC diversity receivers over exponential-generalized Gamma fading model with $E_s/N_0=10$ dB. Solid line: $L=2$ and dashed line: $L=4$.}
\label{fig06}
\end{figure}
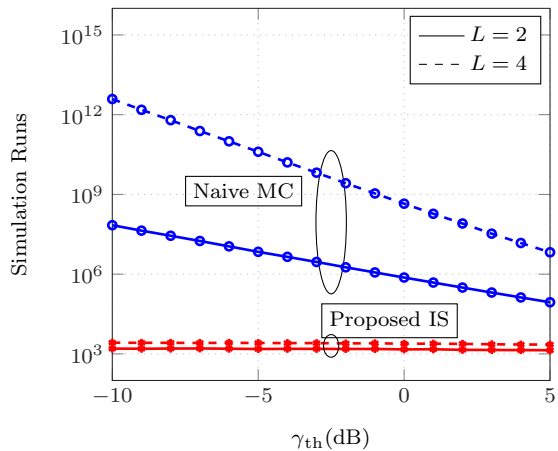

To measure the efficiency of the proposed IS estimator compared to the naive MC estimator, we need to compare the performance of both estimators for a fixed accuracy requirement, i.e. the number of simulation runs required by each estimator when they achieve the same relative error. It can be shown that, for a fixed accuracy requirement $\epsilon_0$, the number of samples needed by naive MC simulations and IS are respectively given by \cite{CO}
\begin{align}
N &= P (1-P) \left(\frac{C}{P \epsilon_0}\right)^2,\\
N^{*} &= \mathbb{V}^*\left[\mathbbm{1}_{\left( S_L(\mathbf{X}) \leq \gamma_0 \right)} \frac{f(\mathbf{X})}{f^*(\mathbf{X})}\right] \left(\frac{C}{P \epsilon_0}\right)^2,
\end{align}
where the constant $C = 1.96$ corresponds to a $95\%$ confidence interval of both estimators, and we distinguish between the number of simulation runs of naive MC, $N$, and the one of IS, $N^*$.\\
In Fig. \ref{fig02} (respectively Fig. \ref{fig06}), we plot the number of simulations runs required by naive MC and IS to achieve a $5\%$ accuracy level in the presence of the exponential-lognormal (respectively the exponential-generalized Gamma) fading model. We clearly see that in both plots, unlike naive MC where the number of samples is rapidly increasing as the outage probability becomes smaller, the proposed IS scheme seems to use an almost constant number of samples no matter how small the outage probability is. Although we do not prove such statement here, we conjecture that the proposed IS estimator is endowed with the bounded relative error \cite{CO}. This in particular will mean that no matter how small the outage probability, the number of simulation runs required by the proposed IS scheme will remain bounded unlike naive MC where the number of samples increases as the probability becomes smaller.

\section{Conclusion}
In this paper, we presented an IS estimator approach to evaluate the outage probability of maximum ratio combining receivers over exponential-lognormal or exponential-generalized Gamma fadings in underwater wireless optical channels. Simulation results show that the proposed approach results in significant computational savings as the outage probability becomes smaller.

\nocite{*}
\bibliographystyle{elsarticle-num}
\bibliography{reference}
  
\end{document}